# Field-tunable quantum disordered ground state in the triangular lattice antiferromagnet NaYbO$_2$

Mitchell Bordelon[1], Eric Kenney[2], Chunxiao Liu[3], Tom Hogan[4], Lorenzo Posthuma[1], Marzieh Kavand[5], Yuanqi Lyu[5], Mark Sherwin[5], N. P. Butch[6], Craig Brown[6,7], M. J. Graf[2], Leon Balents[8], Stephen D. Wilson[1*]

[1] Materials Department, University of California, Santa Barbara, California 93106, USA
[2] Department of Physics, Boston College, Chestnut Hill, Massachusetts 02467, USA
[3] Department of Physics, University of California, Santa Barbara, California 93106, USA
[4] Quantum Design, Inc., San Diego, California 92121, USA
[5] Department of Physics and Center for Terahertz Science and Technology, University of California, Santa Barbara, California 93106, USA
[6] NIST Center for Neutron Research, National Institute of Standards and Technology, Gaithersburg, Maryland 20899, USA
[7] Chemical and Biomolecular Engineering, University of Delaware, Newark, Delaware 19716, USA
[8] Kavli Institute for Theoretical Physics, University of California, Santa Barbara, Santa Barbara, California 93106, USA

* stephendwilson@ucsb.edu

**Antiferromagnetically coupled S=1/2 spins on an isotropic triangular lattice is the paradigm of frustrated quantum magnetism, but structurally ideal realizations are rare. Here we investigate NaYbO$_2$, which hosts an ideal triangular lattice of J$_{eff}$=1/2 moments with no inherent site disorder. No signatures of conventional magnetic order appear down to 50 mK, strongly suggesting a quantum spin liquid ground state. We observe a two-peak specific heat and a nearly quadratic temperature dependence in accord with expectations for a two-dimensional Dirac spin liquid. Application of a magnetic field strongly perturbs the quantum disordered ground state and induces a clear transition into a collinear ordered state consistent with a long-predicted "up-up-down" structure for a triangular lattice XXZ Hamiltonian driven by quantum fluctuations. The observation of spin liquid signatures in zero field and quantum-induced ordering in intermediate fields in the same compound demonstrate an intrinsically quantum disordered ground state. We conclude that NaYbO$_2$ is a model, versatile platform for exploring spin liquid physics with full tunability of field and temperature.**

Exotic ground states of quantum antiferromagnets are encouraged by the combination of low dimensionality, geometric frustration, and inherent anisotropies. Planar triangular lattices are long sought platforms for stabilizing them[1–7]; however, ideal manifestations that do not break crystallographic or exchange symmetries upon approaching the quantum regime are rare. The organic compounds κ-(BEDT-TTF)$_2$Cu$_2$(CN)$_3$ [8] and EtMe$_3$Sb[Pd(dmit)$_2$]$_2$ [9] are two promising examples of triangular lattices with $S$=1/2 moments and a dynamically disordered spin ground state. However, $S$=1/2 inorganic analogs such as Ba$_3$CoSb$_2$O$_9$ [10,11] and NaTiO$_2$ [12–14] either order magnetically or undergo a lattice deformation upon cooling. A roadblock in inorganic systems is the identification of a material with a high crystallographic symmetry, rigid structure, and minimal defect mechanisms that also contains magnetic ions possessing strong quantum fluctuations. Ideally, the magnetic ions should reside at high symmetry positions that preclude antisymmetric Dzyaloshinskii-Moriya exchange from lifting geometric frustration.

As an alternative to $S$=1/2 based compounds, rare earth ions with ground state doublets also engender enhanced quantum fluctuations. Recent studies have shown that the spin-orbit entangled $J_{eff}$=1/2 moments of Yb$^{3+}$ ions on a triangular lattice may exhibit a variety of nearly degenerate states[15–22]. Given the appropriate anisotropies and proximity to phase boundaries, spin liquid states are predicted to appear[22]. Recent experimental studies of one candidate material YbMgGaO4 report continuum spin dynamics and a spin liquid-like ground state, but chemical disorder endemic to the mixed occupancies of Mg and Ga atoms on the same crystallographic site of this material has clouded interpretation[15–23]. Alternate models propose that the random cation distribution incites exchange disorder, facilitating the formation of a weakly-bound spin glass state that experimentally resembles a spin liquid[24–26]. Fully removing this exchange disorder and accessing the physics inherent to an ideal triangular lattice of $J_{eff}=1/2$ moments remains an outstanding challenge.

The rare earth moments in compounds of the form NaRO2 (R=rare earth ions) are known to form an ideal triangular lattice in the $\alpha$-NaFeO2 (*R*-3m) structure. Previous studies suggest that they realize a large degree of magnetic frustration [27–29], and NaYbO2 in particular stands out as an appealing candidate material. Specifically, the NaYbO2 lattice promotes enhanced exchange through short nearest neighbor bonds, and the Yb moments occupy high-symmetry sites that forbid Dzyaloshinskii-Moriya interactions. This combined with large crystal field splitting between the ground state and first excited doublet[20,21] render this lattice an appealing framework.

Here we present an investigation of the triangular lattice of Yb moments in NaYbO2. Our data show that the lattice is robust and forms with fully occupied Na and O sites, minimizing potential chemical/exchange disorder. Furthermore, the structure preserves the ideal equilateral triangular lattice into the quantum regime beyond the point in which the majority of the magnetic entropy has frozen out. Zero field susceptibility data collected down to 50 mK reveal no signatures of spin freezing or glassiness, and heat capacity data collected over the same temperature range reveal only a broad two-peak structure—a common signature of the onset of short-range correlations in materials thought to host quantum disordered ground states[30–33]. This disordered state is strongly perturbed via the application of a magnetic field that induces antiferromagnetic order consistent with an *up-up-down* plateau state for the triangular lattice and reflects an underlying XXZ Hamiltonian with enhanced fluctuations due to interlayer frustration. Our data reveal that NaYbO2 hosts an enticing quantum disordered ground state controllable via modest external fields and presents a cleanly tunable platform for exploring order to quantum disorder phase boundaries inherent to the XXZ triangular lattice.

Polycrystalline NaYbO2 was synthesized and characterized via neutron powder diffraction measurements (see Methods section for further synthesis and experiment details). Figure 1 (a) shows the structure at 1.6K, revealing *R-3m* symmetry with fully occupied Na and O sites. The $D_{3d}$ distorted YbO6 octahedra and bond lengths are illustrated, and a similar YbO6 environment in YbMgGaO4 is known to generate a large 38meV splitting between the first excited state and the ground state doublet[20,21]. A comparison of the Yb environments of the two materials is provided in Supplemental Table 2. At low temperatures, the ground state therefore behaves as an isolated $J_{eff}=1/2$ Kramers doublet. Nearest-neighbor Yb-Yb distances were refined to 3.3507(1) Å at 300 K, consistent with previous reports[27–29], and naively support enhanced exchange relative to other frustrated Yb-based compounds.

Characterizing this exchange, magnetic susceptibility and magnetization data are plotted in Figs. 2 and 3. Below 50 K, the Van Vleck contribution to the susceptibility is negligible. The data from 20K – 100K were modeled by Curie-Weiss fits of the form $\frac{1}{\chi-\chi_0} = \left(\frac{C}{T-\theta_{CW}}\right)^{-1}$ shown in Fig. 2 (a) and yield a local moment of 2.63(8) $\mu_B$ with an antiferromagnetic Curie-Weiss $\theta_{CW} = -10.3(8)$ K. $\theta_{CW}$ is substantially enhanced relative to YbMgGaO$_4$ ($\theta_{CW} = -4$ K [15,16,18,20]), consistent with enhanced exchange. The local moment suggests a heightened *g*-factor, which was validated via electron paramagnetic resonance (EPR) measurements (Fig. 2 (a) inset). A powder averaged *g*-factor of $g_{avg}$=3.03 is implied by the local moment extracted from susceptibility data. Fits to the EPR line shape reveal an anisotropic *g*-factor of $g_{ab}$ = 3.294(8) and $g_c$ = 1.726(9), corresponding to the triangular *ab*-plane and *c*-axis respectively.

While the saturated moment for this system is expected to be approximately 1.6 μ$_B$/Yb, *M(H)* measurements collected at 2 K up to 9 T (Fig. 2 (b)) were only able to polarize Yb moments up to 1 μ$_B$, consistent with significant $\theta_{CW}$ exchange. Fig. 2 (c) shows zero-field AC susceptibility data collected down to 50 mK. No signatures of freezing, frequency dependence, or long-range order are observed. Instead, χ´(*T*) continues to diverge as the sample is cooled, generating an empirical frprintustration parameter $\theta_{CW}$/T$_{AF}$ > 500. As we will later argue, this zero-field state is an inherently quantum disordered state dressed by a small fraction of free Yb moments that are quenched in a magnetic field.

$\chi'(T)$ data collected under a variety of *H* fields are plotted in Fig. 2 (d). Under small *H*, the divergence in $\chi'(T)$ is suppressed and a maximum appears. The temperature of this maximum increases with field until μ$_0$*H*=2T is reached, beyond which $\chi'(T)$ becomes nearly temperature independent. The inflection in $\chi'(T)$ increases linearly with *H* and is plotted in Fig. 1 (b). This matches the expected Zeeman splitting of isolated $J_{eff}$=1/2 moments ΔE=2μ$_B$g$_{avg}J_{eff}$H and suggests that $\chi'(T)$ at μ$_0$*H*=2T represents the remaining majority of the correlated/bound Yb moments. As an estimate of the fraction of free spins, μ$_0$*H*=0T data were fit to a Curie-Weiss form after removing the majority response accessed at μ$_0$*H*=2T. Fits to a Curie-Weiss form between 1 – 4 K (Fig. 2 (d)) are described by a model of 14.4(6)% free spins with a full moment of 2.63 μ$_B$ and a $\theta_{CW} = -0.45(4)$ K. 2K *M(H)* data plotted in Supplementary Fig 2 (d) were also fit to a two-component model of Brillouin-like free spins and exchange-field-bound moments which yielded a free spin fraction of approximately 7%. These fits roughly parametrize the limits of a free spin fraction in the material and suggest that free spins coexist within a quantum disordered ground state.

At higher fields, the nearly temperature-independent $\chi'(T)$ at 2T evolves into an ordered state. Isothermal $\chi'(H)$ data at 330 mK plotted in Fig. 3 (a) show an increase in the susceptibility as a phase boundary is traversed at 3T followed by near total suppression of $\chi'(H)$ at 5T. For μ$_0$*H*>5T, $\chi'(H)$ begins to recover suggesting a higher field phase boundary—one marking the quenching of the ordered state as spins are further polarized toward a quantum paramagnetic phase. $\chi'(T)$ data collected across the ordered regime are plotted in Fig. 3 (b) and show a sharp transition below 1K at μ$_0$*H*=4T. The likely origin of the enhancement in $\chi'(T)$ upon entering the ordered state at 3T is the proximity of the quantum critical point associated with the nearby 0K phase boundary. These quantum fluctuations are suppressed crossing the finite temperature phase boundaries away from this point.

To further characterize NaYbO$_2$, heat capacity measurements were performed. Fig. 3 (c) shows the zero-field C(*T*) of both NaYbO$_2$ and a nonmagnetic comparator NaLuO$_2$ plotted from 80 mK to 40 K. Consistent with susceptibility data, no sharp anomaly indicative of the onset of long-range order is observed in NaYbO$_2$. Instead, a broad feature comprised of two peaks is apparent—one peak centered near 1K and the other near 2.5K. Two peaks in C(*T*) are predicted in a number of theoretical models for both triangular[34,35] and kagome-based[36,37] Heisenberg lattices where a quantum spin liquid state appears. Integrating S$_{mag}$(*T*) data with the lattice contribution subtracted yields a magnetic entropy reaching 95% of $Rln(2)$, consistent with the nominal $J_{eff}$=1/2 magnetic doublet of NaYbO$_2$.

Upon applying magnetic field, data in Fig. 3 (d) show the 2.5 K peak in $C_p(T)$ shifts upward in temperature similar to other frustrated magnets; however, under 5T, a sharp anomaly appears near 1 K and is coincident with the downturn in $\chi'(T)$ at this field. Under 9T, this sharp peak broadens and shifts lower in temperature as the system is driven into the disordered state. $S_{mag}(T)$ integrated under 5T matches that of 0T and the lowest temperature $C_p(T)$ is strongly suppressed once order is generated. This is consistent with the suppression of low energy spin fluctuations upon entering the ordered state, which return when the high field quantum paramagnetic phase is approached. Determining the precise form of the zero-field $C_p(T)$ is complicated by a nuclear Schottky feature that dominates below 100 mK (Supplemental Fig. 2 (c)); however attempts to do so away from this feature yield a $C(T) \propto T^2$ as shown in the inset of Fig. 3 (d).

Low temperature neutron scattering measurements were also performed. Fig. 4 (a) shows temperature subtracted (330mK-1.6K) diffraction data and the absence of zero-field magnetic order. Field subtracted data at 450mK plotted in Fig. 4 (b) reveal that under 5T, new superlattice reflections appear at the **Q**=(1/3, 1/3, 0), (1/3, 1/3, 1), and (1/3, 1/3, 3) positions. Given the symmetry constraints of the *R-3m*, structure, these either represent a 120° noncollinear spin structure or an *up-up-down* pattern of spin order. The absence of a reflection at **Q**=(1/3, 1/3, 2) suggests the field-induced order is collinear. Additionally, magnetic intensity appears at the **Q**=(0, 0, 3) position, consistent with the two-**q** structure (**q**=(1/3, 1/3, 0) + **q**=(0, 0, 0)) expected for the equal moment *up-up-down* state[38]. The best fit to this model is shown in Fig. 4 (c) where spins refine to be oriented nearly parallel to the (1,-1,-1) direction with an ordered moment 1.36±0.1 $\mu_B$. This value is less than the 1.6 $\mu_B$ expected likely due to the presence of a minority fraction of free moments as well as the influence of remnant fluctuations in the ordered state. Magnetic peaks are resolution-limited with a minimum spin-spin correlation length of $\xi_{min}$=450 Å. Further data collected at 67 mK determine the low temperature, magnetic field phase boundaries (Supplementary Fig. 4 (b)).

Inelastic scattering data plotted in Figs. 4 (d) and (e) reveal a renormalization of the low energy spin dynamics upon transitioning from the quantum disordered state into the *up-up-down* phase. The zero field data shows a diffuse spectrum of excitations centered about the (1/3, 1/3, *L*)-type wave vectors, and upon applying a 5T field, much of this spectral weight is shifted into the elastic channel and a nearly flat band of excitations centered at 1 meV. Powder averaged linear spin wave calculations assuming a purely two-dimensional triangular lattice in a 5T magnetic field reproduce this flat feature, and the simulated $S(Q,\omega)$ is plotted in Fig. 4 (f). This simulation was generated using nearest neighbor coupling with a nearly Heisenberg Hamiltonian with a slight easy-plane anisotropy, $J_z$=0.45, $J_{xy}$=0.51 meV. The subtle downturn at low **Q** of the 1 meV band requires easy-plane anisotropy as discussed in the supplemental material.

We now discuss the implications of our results. The similar $YbO_6$ octahedra of $NaYbO_2$ and $YbMgGaO_4$ intimate that the local crystal fields and in-plane exchange couplings between Yb ions are comparable; however, the main distinction between the two systems is the much shorter inter-plane distance in $NaYbO_2$. This suggests that the interlayer coupling is non-negligible, and therefore, a minimal Hamiltonian should include nearest-neighbor bonds within the planes and between neighboring layers. Based on the structure, a symmetry analysis leads to the following exchange Hamiltonian:

$$H_{2d} = \sum_{\langle ij \rangle} \{J_{xy}(S_i^x S_j^x + S_i^y S_j^y) + J_z S_i^z S_j^z + J_c(\hat{e}_{ij} \cdot S_i)(\hat{e}_{ij} \cdot S_j) + J_{cz}[(\hat{z} \cdot \hat{e}_{ij} \times S_i)S_j^z + (\hat{z} \cdot \hat{e}_{ij} \times S_j)S_i^z]\}$$

$$H' = \sum_{\langle\langle ij\rangle\rangle} \{J'_{xy}(S_i^x S_j^x + S_i^y S_j^y) + J'_z S_i^z S_j^z + J_c(\hat{f}_{ij}\cdot S_i)(\hat{f}_{ij}\cdot S_j) + J'_{cz}[(\hat{f}_{ij}\cdot S_i)S_j^z + (\hat{f}_{ij}\cdot S_j)S_i^z]\}$$

The first line contains interactions within a triangular layer, and the second between layers. The unit vectors $e_{ij}$ are oriented along the ij bond, and $f_{ij}$ is a unit vector along the projection of the ij bond into the ab-plane. The in-plane Hamiltonian is identical to that in YbMgGaO$_4$ but rewritten here (following Iaconis et al.[39]) in a more physically transparent "compass model" form. The interlayer exchange in the second line also has a compass-like structure. We expect that this form applies to the full family of delafossite-like antiferromagnets, ARX$_2$, with dipolar Kramer's doublets on the R site, sharing the space group 166.

$H_{2d}$ notably contains a wide range of phase space favoring three types of classical orders: (1) three-sublattice 120° structures; (2) collinear two-sublattice stripe phases; and (3) out-of-plane Ising anisotropy with up-up-down structures. Because we do not observe zero-field order, and we expect that interplane exchange is substantial, we infer that the interactions in *H'* should be frustrated by the in-plane order or correlations. Consideration of the coupling between layers uniquely singles out the three-sublattice 120° structure: to leading order, only this in-plane order allows the staggered magnetization to effectively cancel the exchange field between neighboring planes (see supplementary material)[39]. This is even true to a large extent also for the anisotropic J'$_c$ and J'$_{cz}$ couplings.

Using this deduction that NaYbO$_2$ has 120° correlations, we expect fluctuations amongst many classically degenerate or nearly degenerate states to strongly suppress order. Furthermore, recent DMRG studies of $H_{2d}$ find that for S=1/2 quantum spins, a spin liquid state indeed occurs in a corner of the classically 120° ordered phase space with moderate J$_{cz}$ coupling[40]. Consequently, it is plausible that a spin liquid state occurs in NaYbO$_2$, and if so, it is likely to be smoothly connected to the spin liquid of the two-dimensional problem. The optimal spin liquid ground state for the 2d model based on variational parton calculations[39] is a U(1) Dirac state with gapless fermionic spinons described theoretically as a 2+1-dimensional conformal field theory: QED$_3$. The second implication of our Hamiltonian in this regime is that, on applying a magnetic field, the degeneracy is strongly lifted. This is because a large part of the zero-field cancellation is reliant on the specific 120° structure of the in-plane ground state, which is modified by the application of a magnetic field. Therefore, it is natural to expect ordering to become more robust in an applied magnetic field. The three-sublattice (1/3,1/3,0) wave vector is indeed germane to triangular antiferromagnets in a magnetic field, which stabilize a quantized magnetization plateau at 1/3 saturation in XXZ models[41,42].

With this in mind, we return to a discussion of the data. Theory predicts the 2d U(1) Dirac state to have C$_p(T)$ quadratic in temperature, consistent with measurements[43]. An alternate explanation of $T^2$ specific heat might come from the degenerate line of spiral states found by Rastelli and Tassi for the zero-field rhombohedral XXZ model[44], which has 2d-like spin fluctuations despite 3d coupling. The incommensurate long-range order of the Rastelli-Tassi spiral does not appear in our measurements, however, the field-induced Bragg peaks seen in experiment are consistent with the three-sublattice plateau states that emerge in the XXZ model in a field[41,45]. Indeed, the magnetization at 5T, where the ordered phase is maximal, is approximately 1/3 of the expected saturation moment and corresponds to a plateau where χ(T)=∂M/∂H reaches zero. The best fit to neutron diffraction data further corresponds to the equal moment, two-**q** *up-up-down* structure of the plateau state.

The two peaks observed in the zero field $C_p(T)$ of NaYbO$_2$ evoke a number of theoretical models of Heisenberg spins on both triangular[34,35] and kagome[36,37] lattices that predict dual entropy anomalies upon cooling into spin liquid ground states. Both peaks are rarely observed experimentally and interpretations of the nature of each peak vary with the specific model. Exact diagonalization studies of the XXZ Hamiltonian on a triangular lattice predict a high temperature peak corresponding to the formation of trimers of doublet states (i.e. short-range correlations) followed by a lower temperature peak that marks the onset of a quantum spin singlet state[35]. Recent work exploring the $S=1/2$ triangular lattice using tensor renormalization group techniques predicts a dual $C_p(T)$ anomaly with the lower temperature peak signifying the onset of short-range/incipient order and the upper peak reflective of the onset of gapped low energy, chiral fluctuations[31]. The ratio of peak temperatures predicted in this $S=1/2$ model $T_l/T_h \approx 0.36$ is consistent with those observed in NaYbO$_2$ and the $J \approx 5$ K inferred from the model is reasonably close to the $\theta_{CW}$ determined from susceptibility data.

Our data demonstrate that the nearly ideal triangular lattice of Yb ions in NaYbO$_2$ realize an unconventional quantum disordered ground state. Unlike the majority of other spin-liquid candidates such as Herbertsmithite ZnCu$_3$(OH)$_6$Cl$_2$ [47], the ground state in NaYbO$_2$ can be driven into an intermediate antiferromagnetic ordered regime in relatively weak magnetic fields. The origin of the small fraction of free spins coexisting with this ground state remains an open question; however, they are not reflective of trivial disorder which favors the least collinear state[46]. Additionally, rather than hosting a purely two-dimensional network of spins where the two-dimensionality precludes long-range order such as in Ba$_8$CoNb$_6$O$_{24}$[48,49], interlayer geometric frustration is critical to the exclusion of order in NaYbO$_2$. This reflects the strong perturbation field provides to a complex interplay between interlayer frustration and the nearly degenerate ground states inherent to the XXZ triangular lattice Hamiltonian. Due to this, NaYbO$_2$ uniquely stands able to provide considerable insight into the critical phase behavior manifest at the phase boundaries between the ordered and quantum disordered states in a chemically-ideal frustrated triangular lattice.

**References:**

(1) P.W. Anderson. Resonating valence bonds: a new kind of insulator? *Mater. Res. Bull.* **8**, 153-160 (1973).
(2) P.W. Anderson. The Resonating Valence Bond State in La2CuO4 and Superconductivity. *Science* **235**, 1196-1198 (1987).
(3) P.A. Lee. An end to the drought of quantum spin liquids. *Science* **321**, 1306-1307 (2008).
(4) L. Balents. Spin liquids in frustrated magnets. *Nature* **464**, 199-208 (2010).
(5) L. Savary & L. Balents. Quantum spin liquids: a review. *Rep. Prog. Phys.* **80**, 016502 (2017).
(6) Witczak-Krempa, W.; Chen, G.; Kim, Y.B.; Balents, L. Correlated Quantum Phenomena in the Strong Spin-Orbit Regime. *Annu. Rev. Condens. Matter Phys.* **5**, 57-82 (2014).
(7) Zhou, Y.; Kanoda, K.; Ng, T.-K. Quantum Spin Liquid States. *Rev. Mod. Phys.* **89**, 025003 (2017).
(8) S.-S. Lee & P.A. Lee. U(1) Gauge Theory of the Hubbard Model: Spin Liquid States and Possible Application to κ-(BEDT-TTF)2Cu2(CN)3. *Phys. Rev. Lett*. **95**, 036403 (2005).
(9) Itou, T.; Oyamada, A.; Maegawa, S.; Tamura, M.; Kato, R. Quantum spin liquid in the spin-1/2 triangular antiferromagnet EtMe3Sb[Pd(dmit)2]2. *Phys. Rev. B* **77**, 104413 (2008).
(10) Ma, J.; Kamiya, Y.; Hong, T.; Cao, H.B.; Ehlers, G.; Tian, W.; Batista, C.D.; Dun, Z.L.; Zhou, H.D.; Matsuda, M. Static and Dynamical Properties of the Spin-1/2 Equilateral Triangular-Lattice Antiferromagnet Ba3CoSb2O9. *Phys. Rev. Lett.* **116**, 087201 (2016).

(11) Shirata, Y.; Tanaka, H.; Matsuo, A.; Kindo, K. Experimental realization of a spin-1/2 triangular-lattice Heisenberg antiferromagnet. *Phys. Rev. Lett.* **108**, 057205 (2012).

(12) G. Jackel & D.A. Ivanov. Dimer phases in quantum antiferromagnets with orbital degeneracy. *Phys. Rev. B* **76**, 132407 (2007).

(13) Clarke, S.J.; Fowkes, A.J.; Harrison, A.; Ibberson, R.M.; Rosseinsky, M.J. Synthesis, Structure, and Magnetic Properties of NaTiO2. *Chem. Mater.* **10**, 372-384 (1998).

(14) McQueen, T.M.; Stephens, P.W.; Huang, Q.; Klimczuk, T.; Ronning, F.; Cava, R.J. Successive Orbital Ordering Transitions in NaVO2. *Phys. Rev. Lett.* **101**, 166402 (2008).

(15) Li, Y.; Liao, H.; Zhang, Z.; Li, S.; Jin, F.; Ling, L.; Zhang, L.; Zou, Y.; Pi, L.; Yang, Z.; Wang, J.; Wu, Z.; Zhang, Q. Gapless quantum spin liquid ground state in the two-dimensional spin-1/2 triangular antiferromagnet YbMgGaO4. *Sci. Rep*. **5**, 16419 (2015).

(16) Li, Y.; Chen, G.; Tong, W.; Pi, L.; Liu. J.; Yang, Z.; Wang, X.; Zhang, Q. Rare-Earth Triangular Lattice Spin Liquid: A Single-Crystal Study of YbMgGaO4. *Phys. Rev. Lett*. **115**, 167203 (2015).

(17) Li, Y.; Adroja, D.; Biswas, P.K.; Baker, P.J.; Zhang, Q.; Liu, J.; Tsirlin, A.A.; Gegenwart, P.; Zhang, Q. Muon Spin Relaxation Evidence for the U(1) Quantum Spin-Liquid Ground State in the Triangular Antiferromagnet YbMgGaO4. *Phys. Rev. Lett*. **117**, 097201 (2016).

(18) Shen, Y.; Li, Y.-D.; Wo, H.; Li, Y.; Shen, S.; Pan, B.; Wang, Q.; Walker, H.C.; Steffens, P.; Boehm, M.; Hao, Y.; Quintero-Castro, D.L.; Harriger, L.W.; Frontzek, M.D.; Hao, L.; Meng, S.; Zhang, Q.; Chen, G.; Zhao, J. Evidence for a spinon Fermi surface in a triangular-lattice quantum-spin-liquid candidate. *Nature* **540**, 559-562 (2016).

(19) Xu, Y.; Zhang, J.; Li, Y. S.; Yu, Y. J.; Hong, X. C.; Zhang, Q. M.; Li, S. Y. Absence of Magnetic Thermal Conductivity in the Quantum Spin-Liquid Candidate YbMgGaO4. *Phys. Rev. Lett*. **117**, 267202 (2016).

(20) Paddison, J.A.M.; Daum, M.; Dun, Z.L.; Ehlers, G.; Liu, Y.; Stone, M.B.; Zhou, H.D.; Mourigal, M. Continuous excitations of the triangular-lattice quantum spin liquid YbMgGaO$_4$. *Nat. Phys.* **13**, 117-122 (2017).

(21) Li, Y.; Adroja, D.; Bewley, R.I.; Voneshen, D.; Tsirlin, A.A.; Gegenwart, P.; Zhang, Q. Crystalline Electric-Field Randomness in the Triangular Lattice Spin-Liquid YbMgGaO4. *Phys. Rev. Lett*. **118**, 107202 (2017).

(22) Li, Y.-D.; Wang, X.; Chen, G. Anisotropic spin model of strong spin-orbit-coupled triangular antiferromagnets. *Phys. Rev. B* **94**, 035107 (2016).

(23) Li, Y.-D.; Shen, Y.; Li, Y.; Zhao, J.; Chen, G. Effect of spin-orbit coupling on the effective-spin correlation in YbMgGaO4. *Phys. Rev. B* **97**, 125105 (2018).

(24) Zhu, Z.; Maksimov, P.A.; White, S.R.; Cheryshev, A.L. Disorder-Induced Mimicry of a Spin Liquid in YbMgGaO4. *Phys. Rev. Lett* **119**, 157201 (2017).

(25) Kimchi, I.; Nahum, A.; Senthil, T. Valence Bonds in Random Quantum Magnets: Theory and Application to YbMgGaO4. *Phys. Rev. X* **8**, 031028 (2018).

(26) Zhen Ma, Z.; Jinghui Wang, J.; Dong, Z.-Y.; Zhang, J.; Li, S.; Zheng, S.-H.; Yu, Y.; Wang, W.; Che, L.; Ran, K.; Bao, S.; Cai, Z.; Čermák, P.; Schneidewind, A.; Yano, S.; Gardner, J.S.; Lu, X.; Yu, S.-L.; Liu, J.-M.; Li, S.; Li, J.-X.; Wen, J. Spin-Glass Ground State in a Triangular-Lattice Compound YbZnGaO4. *Phys. Rev. Lett.* **120**, 087201 (2018).

(27) Hashimoto, Y.; Wakeshima, M.; Hinatsu, Y. Magnetic properties of ternary sodium oxides NaLnO2 (Ln=rare earths). *J. Solid State. Chem.* **176**, 266-272 (2003).

(28) Liu, W.; Zhang, Z.; Ji, J.; Liu, Y.; Li, J.; Wang, X.; Lei, H.; Chen, G.; Zhang, Q. Rare-Earth Chalcogenides: A Large Family of Triangular Lattice Spin Liquid Candidates. *Chin. Phys. Lett*. **35**, 117501 (2018).

(29) Baenitz, M.; Schlender, Ph.; Sichelschmidt, J.; Onykiienko, Y.A.; Zangeneh, Z.; Ranjith, K.M.; Sarkar, R.; Hozoi, L.; Walker, H.C.; Orain, J.-C.; Yasuoka, H.; van den Brink, J.; Klauss, H.H.; Inosov, D.S.; Doert, Th. NaYbS2: A planar spin-1/2 triangular-lattice magnet and putative spin liquid. *Phys. Rev. B* **98**, 220409(R) (2018).

(30) C. Zeng & V. Elser. Numerical studies of antiferromagnetism on a Kagomé net. *Phys. Rev. B* **42**, 8436 (1990).

(31) Chen, L.; Qu, D.-W.; Li, H.; Chen, B.-B.; Gong, S.-S.; von Delft, J.; Weichselbaum, A.; Li, W. Two-temperature scales in the triangular-lattice Heisenberg antiferromagnet. *Phys. Rev. B* **99**, 140404(R) (2019).

(32) Nambu, Y.; Nakatsuji, S.; Maeno, Y. Coherent Behavior and Nonmagnetic Impurity Effects of Spin Disordered State in NiGa2S4. *J. Phys. Soc. Jpn.* **75**, 043711 (2006).

(33) Gardner, J.S.; Gingras, M.J.P.; Greedan, J.E. *Rev. Mod. Phys.* **82**, 53 (2010).

(34) Y.R. Wang. Specific heat of a quantum Heisenberg model on a triangular lattice with two exchange parameters and its application to 3He adsorbed on graphite. *Phys. Rev. B* **45**, 12608(R) (1992).

(35) Isoda, M.; Nakano, H.; Sakai, T. Specific Heat and Magnetic Susceptibility of Ising-Like Anisotropic Heisenberg Model on Kagome Lattice. *J. Phys. Soc. Jpn.* **80**, 084704 (2011).

(36) N. Elstner & A.P. Young. Spin-1/2 Heisenberg antiferromagnet on the kagomé lattice: High-temperature expansion and exact-diagonalization studies. *Phys. Rev. B* **50**, 6871-6876 (1994).

(37) R.R.P. Singh & J. Oitmaa. High-temperature series expansion study of the Heisenberg antiferromagnet on the hyperkagome lattice: Comparison with Na4Ir3O8. *Phys. Rev. B* **85**, 104406 (2012).

(38) Garlea, V.O.; Sanjeewa, L.D.; McGuire, M.A.; Batista, C.D.; Samarakoon, A.M.; Graf, D.; Winn, B.; Ye, F.; Hoffmann, C.; Kolis, J.W. Exotic Magnetic Field-Induced Spin-Superstructures in a Mixed Honeycomb-Triangular Lattice System. *Phys. Rev. X* **9**, 011038 (2019).

(39) Iaconis, J.; Liu, C.; Haláz, G.B.; Balents, L. Spin liquid versus spin orbit coupling on the triangular lattice. *SciPost Phys.* **4**, 003 (2018).

(40) Zhu, Z.; Maksimov, P.A.; White, S.R.; Chernyshev, A.L. Topography of Spin Liquids on a Triangular Lattice. *Phys. Rev. Lett.* **120**, 207203 (2018).

(41) O.A. Starykh. Unusual ordered phases of highly frustrated magnets: a review. *Rep. Prog. Phys.* **78**, 052502 (2015).

(42) A. V. Chubokov & D. I. Golosov. Quantum theory of an antiferromagnet on a triangular lattice in a magnetic field. *J. Phys. Condens. Matter* **3**, 69-82 (1991).

(43) Ran, Y.; Hermele, M.; Lee, P.A.; Wen, X.-G. Projected-Wave-Function Study of the Spin-1/2 Heisenberg Model on the Kagomé Lattice. *Phys. Rev. Lett.* **98**, 117205 (2007).

(44) E. Rastelli & A. Tassi. The rhombohedral Heisenberg antiferromagnet: infinite degeneracy of the ground state and magnetic properties of solid oxygen. *J. Phys. C: Solid State Phys.* **19**, L423-L428 (1986).

(45) Ono, T.; Tanaka, H.; Kolomiyets, O.; Mitamura, H.; Goto, T.; Nakajima, K.; Oosawa, A.; Koike, Y.; Kakurai, K.; Klenke, J. Magnetization plateaux of the S= 1/2 two-dimensional frustrated antiferromagnet Cs2CuBr4. *J. Phys. Condens. Matter* **16**, S773-S778 (2004).

(46) V.S. Maryasin & M.E. Zhitomirsky. Triangular Antiferromagnet with Nonmagnetic Impurities. *Phys. Rev. Lett.* **111**, 247201 (2013).

(47) Helton, J.S.; Matan, K.; Shores, M.P.; Nytko, E.A.; Bartlett, B.M.; Yoshida, Y.; Takano, Y.; Suslov, A.; Qui, Y.; Chung, J.-H.; Nocera, D.G. Lee, Y.S. Spin Dynamics of the Spin-1/2 Kagome Lattice Antiferromagnet ZnCu3(OH)6Cl2. *Phys. Rev. Lett.* **98**, 107204 (2007).

(48) Rawl, R.; Ge, L.; Agrawal, H.; Kamiya, Y.; Dela Cruz, C.R.; Butch, N.P.; Sun, X.F.; Lee, M.; Choi, E.S.; Oitmaa, J.; Batista, C.D.; Mourigal, M.; Zhou, H.D.; Ma, J. Ba8CoNb6O24: A spin-1/2 triangular-lattice Heisenberg antiferromagnet in the two-dimensional limit. *Phys. Rev. B* **95**, 060412(R) (2017).

(49) Cui, Y.; Dai, J.; Zhou, P.; Wang, P.S.; Li, T.R.; Song, W.H.; Wang, J.C.; Ma, L.; Zhang, Z.; Li, S.Y.; Luke, G.M.; Normand, B.; Xiang, T.; Yu, W. Mermin-Wagner physics, (H,T) phase diagram, and candidate quantum spin-liquid phase in the spin-1/2 triangular-lattice antiferromagnet Ba8CoNb6O24. *Phys. Rev. Mater.* **2**, 044403 (2018).

**Methods:**

**Sample preparation:** Polycrystalline NaYbO$_2$ powder was prepared by a solid-state reaction of Yb$_2$O$_3$ (99.99%, Alfa Aesar) with Na$_2$CO$_3$ (99.997%, Alfa Aesar) in a 1:1.25 molar ratio and reacted at 1000 °C for 3 days with subsequent regrinding and reheating to 1000°C for another day. A slight excess of Na$_2$CO$_3$ remains in the powder and is required to prevent the formation of magnetic Yb$_2$O$_3$ impurities and stabilize the NaYbO$_2$ phase while reacting. In contrast to previous reports[27–29], attempts to remove this excess Na$_2$CO$_3$ via washing with common solvents (acidic/basic/neutral water, methanol, ethanol, isopropanol) induced NaYbO$_2$ degradation and Yb$_2$O$_3$ reformation. Therefore, the initial reaction was optimized to include minimal nonmagnetic Na$_2$CO$_3$ impurities while maintaining complete reaction of Yb$_2$O$_3$ powder. All subsequent measurements accounted for the known Na$_2$CO$_3$ impurity fractions present in samples. All samples were stored in a dry, inert atmosphere and minimally exposed to atmospheric conditions prior to measurements. Neutron diffraction data show that the Na content of NaYbO$_2$ is stoichiometric within resolution (~1%). While managing alkali content in many compounds can be challenging, the strong preference for Yb$^{3+}$ aids in yielding fully occupied Na sites.

**Neutron Diffraction:** Powder neutron diffraction data was collected on the BT1 spectrometer at the National Institute of Standards and Technology Center for Neutron Research, MD. For zero field measurements, the instrument was equipped with a cryostat capable of reaching 1.6 – 300 K. A 7 T vertical field cryostat with a $^3$He insert was used to collect magnetic field data at 0, 5, and 7 T between 0.33 – 1.5 K. Samples were placed in a vanadium canister for high-temperature measurements and a copper cell for low-temperature measurements. The 300 K data was collected with neutrons of wavelengths 1.5399 Å and 2.0774 Å produced by a Cu(311) and Ge(311) monochromators, respectively. All other temperatures utilized neutrons of wavelength 2.0774 Å for maximum intensity. Rietveld refinement was performed with the FullProf software suite[50] and GSAS/EXPGUI programs[51,52], and fits to the data at 300 K and 1.6 K are shown in Supplementary Figure 1. Additional energy-resolved neutron scattering data were collected on the Disk Chopper Spectrometer (DCS) at the NIST Center for Neutron Research (NCNR) within a 10 T magnetic with a dilution insert. Incident neutrons with 5 Å wavelength were used with the medium resolution chopper setting.

**Magnetic measurements**: Magnetic properties were measured using several different instruments. Isothermal DC magnetization up to 9 T was collected on a Quantum Design Physical Property Measurement System (PPMS) with a vibrating sample magnetometer insert, and low field DC magnetic susceptibility from 2K – 300 K was obtained on a Quantum Design Magnetic Property Measurement System (MPMS3). Isothermal AC susceptibility in applied fields up to 7 T and temperature dependent AC magnetic susceptibility was measured on an AC susceptometer at 711.4 Hz with a 0.1 Oe (7.96 A/m) drive field and a $^3$He insert capable of cooling to 330 mK. All fields collected are displayed in Supplementary Figure 1. Within our AC susceptibility measurements at 711.4 Hz, the sample holder's contribution is nearly linear at this range and this frequency dependent background was subtracted. This was confirmed via frequency sweeps of the empty holder loaded with an equivalent amount of silver paint and thermal-link wire used for the experiment. The background frequency response was then subtracted from the data. Further zero-field, frequency-dependent, AC susceptibility data between 10 Hz – 10 kHz and between 50 mK – 4 K was collected with the AC Susceptibility for Dilution Refrigerator (ACDR) option on a Quantum Design PPMS with a 1 Oe drive field. Measurements of the empty sample holder were performed in this ACDR setup at various excitation amplitudes and frequencies to quantify the total background. Trim coils were then used to cancel (null) the measured frequency dependent background.

Electron paramagnetic resonance (EPR) spectra were recorded at 4.2 K with a Bruker EMXplus EPR spectrometer in the perpendicular operation mode. The observed resonance line shape was modeled with the EasySpin package implemented in Matlab[53]. Broadening of the EPR lineshape was observed and could be fit to a normal distribution of the out of plane $g_c$-factor (FWHM = 0.40(7)) with no resolvable broadening in the in-plane $g_{ab}$-factor.

**Heat capacity measurements:** Specific heat between 80 mK – 300 K was performed for sintered pellets of $NaYbO_2$ under zero-field conditions on a Quantum Design PPMS using the Dilution Refrigerator (DR) insert for temperatures below 1.8K. Data under 2.5T, 5T, and 9 T applied fields were also collected. The magnetic portion of specific heat was determined by subtracting the nonmagnetic structural analogue $NaLuO_2$ measured between 2K – 200 K (this was zero within resolution for temperatures lower than 2 K). Calculated entropy was determined by integrating $C_{mag}/T$ between 80 mK to 40 K.

**Methods References:**

(50) Rodríguez-Carvajal, J. Recent advances in magnetic structure determination by neutron powder diffraction. *Phys. B Condens. Matter* **192**, 55-69 (1993).
(51) Larson, A. C. & Von Dreele, R. B. General structure analysis system (GSAS), Los Alamos National Laboratory Report LAUR 86-748 (2000).
(52) Toby, B. H. EXPGUI, A Graphical User Interface for GSAS. *J. Appl.Crystallogr*. **34**, 210-213 (2001).
(53) S. Stoll & A. Schweiger. EasySpin, a comprehensive software package for spectral simulation and analysis in EPR. *J. Magn. Reson.* **178**, 42-55 (2006).
**Data Availability:** The data that support the findings of this study are available from the corresponding author upon reasonable request. Neutron data were collected on the BT-1 diffractometer and the DCS spectrometer at the NIST Center for Neutron Research.

**Acknowledgments:** This work was supported by DOE, Office of Science, Basic Energy Sciences under Award DE-SC0017752 (S.D.W. and M.B.). M.B. acknowledges partial support by the National Science Foundation Graduate Research Fellowship Program under Grant No. 1650114 (M.B.). Work by L.B. and C.L. was supported by the DOE, Office of Science, Basic Energy Sciences under Award No. DE-FG02-08ER46524. Identification of commercial equipment does not imply recommendation or endorsement by NIST.

**Author Contributions**: M.B., S.D.W., C.L., and L.B. wrote the manuscript. M.B., S.D.W. analyzed experiment data and planned experiments. M.J.G. and E.K. performed susceptibility measurements. M.B. and T.H. performed heat capacity and magnetization measurements. C.L. and L.B. performed theoretical analysis of the material. C.B. performed the neutron diffraction measurements and M.B. and N.P.B. performed inelastic neutron scattering experiments. M.S., M.K., Y.L., and M.B. performed electron spin resonance measurements. M.B. and L.P. synthesized the materials.

**Additional Information:** Supplementary Information accompanies this paper.

**Competing interests:** The authors declare no competing financial or non-financial interests.

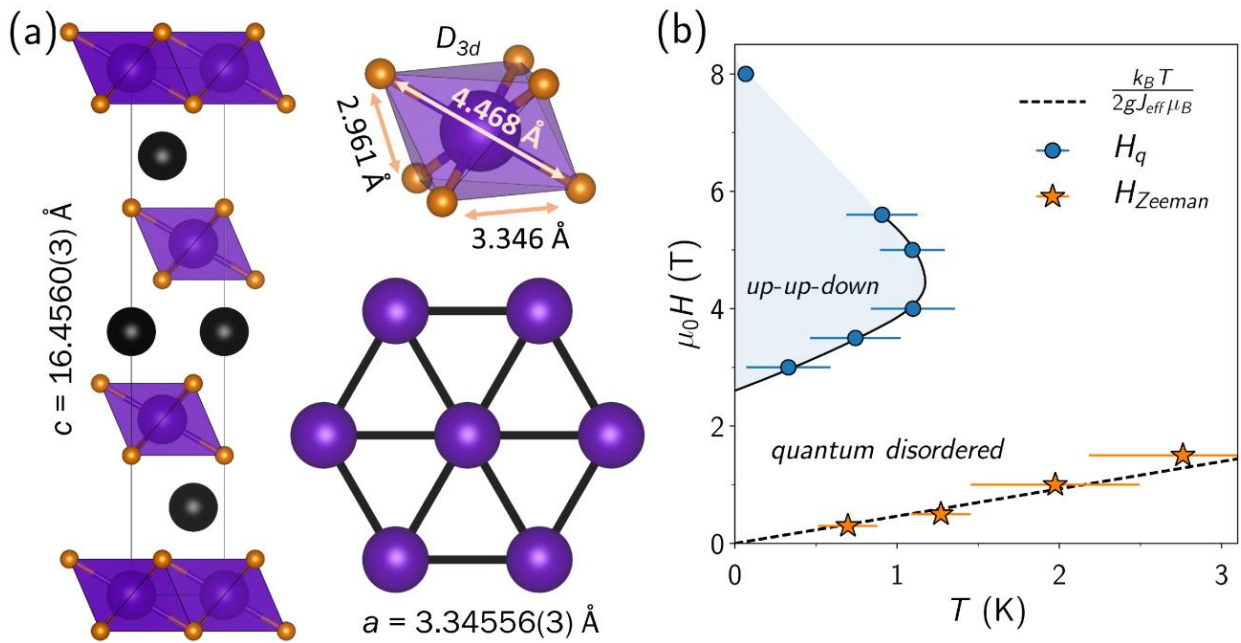

**Figure 1: Crystal structure and magnetic ($H,T$) phase diagram of NaYbO$_2$.** a) Refined NaYbO$_2$ structure (1.6 K, $R$-$3m$) contains equilateral triangular layers of $D_{3d}$ YbO$_6$ distorted octahedra separated by 3.346 Å. Na cations refine to full occupation, creating a uniform chemical environment surrounding the triangular layers. b) Low-temperature phase boundary between quantum disordered (QD) and antiferromagnetic order states in NaYbO$_2$ plotted as a function of field and temperature extracted from AC susceptibility and neutron scattering experiments. Dashed line denotes the boundary of Zeeman-driven quenching of a minority fraction of free Yb moments under field, above which free moments are quenched. These free moments coexist with a quantum disordered ground state. Values in parentheses and error bars indicate one standard deviation.

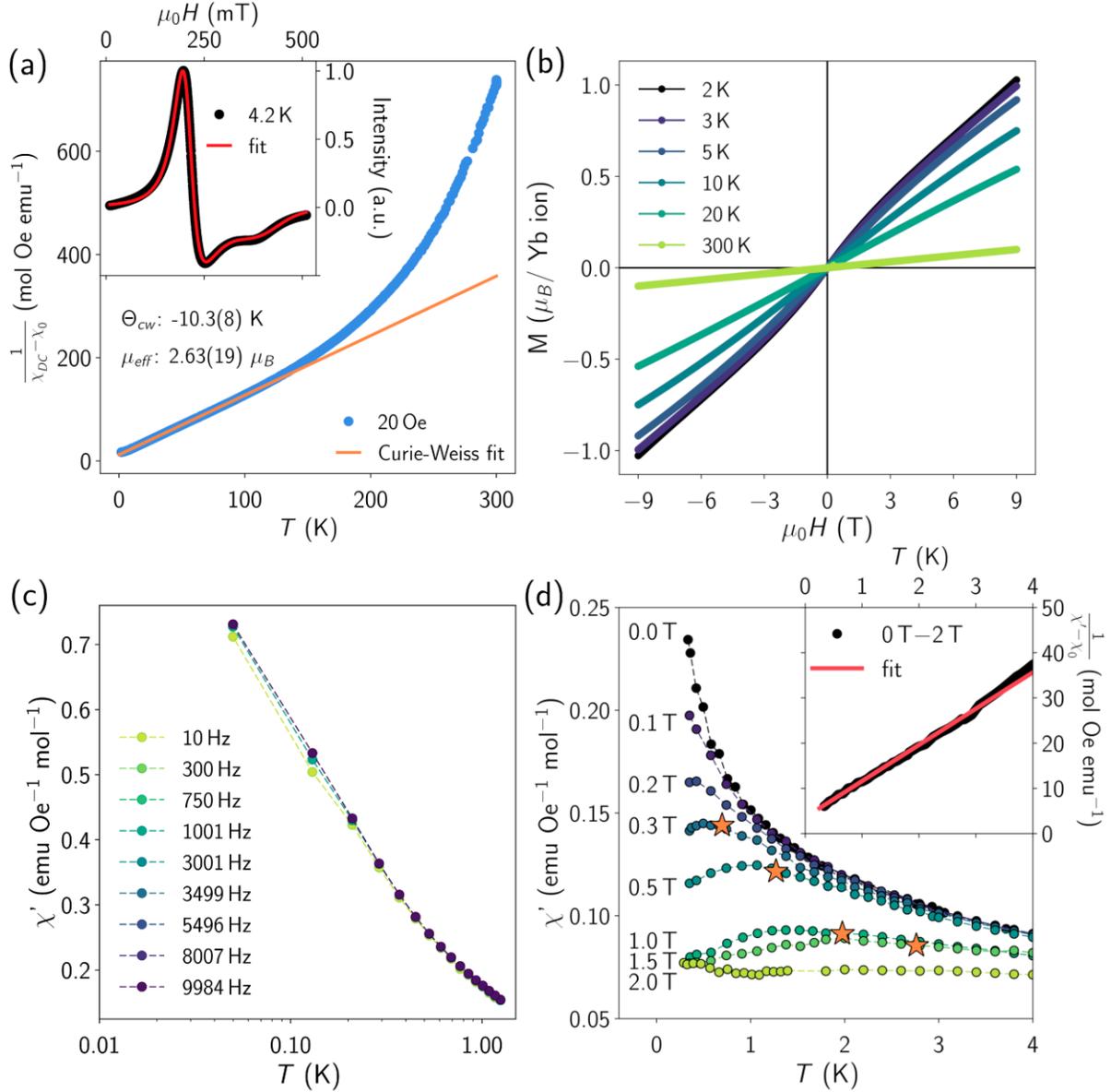

**Fig. 2: Low-field magnetization and magnetic susceptibility data.** (a) Low-temperature Curie-Weiss fit in a temperature range free from Van Vleck contributions from high energy crystal field doublets (*ie*. where the majority of trivalent Yb ions are in the $J_{eff}=1/2$ ground state). A large mean-field interaction strength of -10.3(8) K with an effective local moment value of 2.63(19) $\mu_B$ is fit with a temperature-independent $\chi_0 = 0.0053(3)$ emu mol$^{-1}$ background term. Inset: Electron paramagnetic resonance data collected at 4.2 K fit to anisotropic *g*-factors of $g_{ab} = 3.294(8)$ and $g_c = 1.726(9)$. (b) Isothermal magnetization versus field data reaching only 67% of the expected 1.5 $\mu_B$/Yb-ion polarized moment under $\mu_0H$=9T. (c) Temperature and frequency dependence of AC magnetic susceptibility $\chi'(T)$ from 50 mK to 4 K under zero-field. (d) $\chi'(T)$ data collected under applied magnetic fields. A minority fraction of free Yb moments are quenched at low temperatures and high fields, resulting in a peak in $\chi'(T)$, and the downward inflection parameterizing this Zeeman splitting is denoted by orange stars. Inset shows field subtracted 0T-2T $\chi'(T)$ data between 1K – 3 K and a Curie-Weiss fit quantifying the fraction of free Yb moments in the system as described in the text.

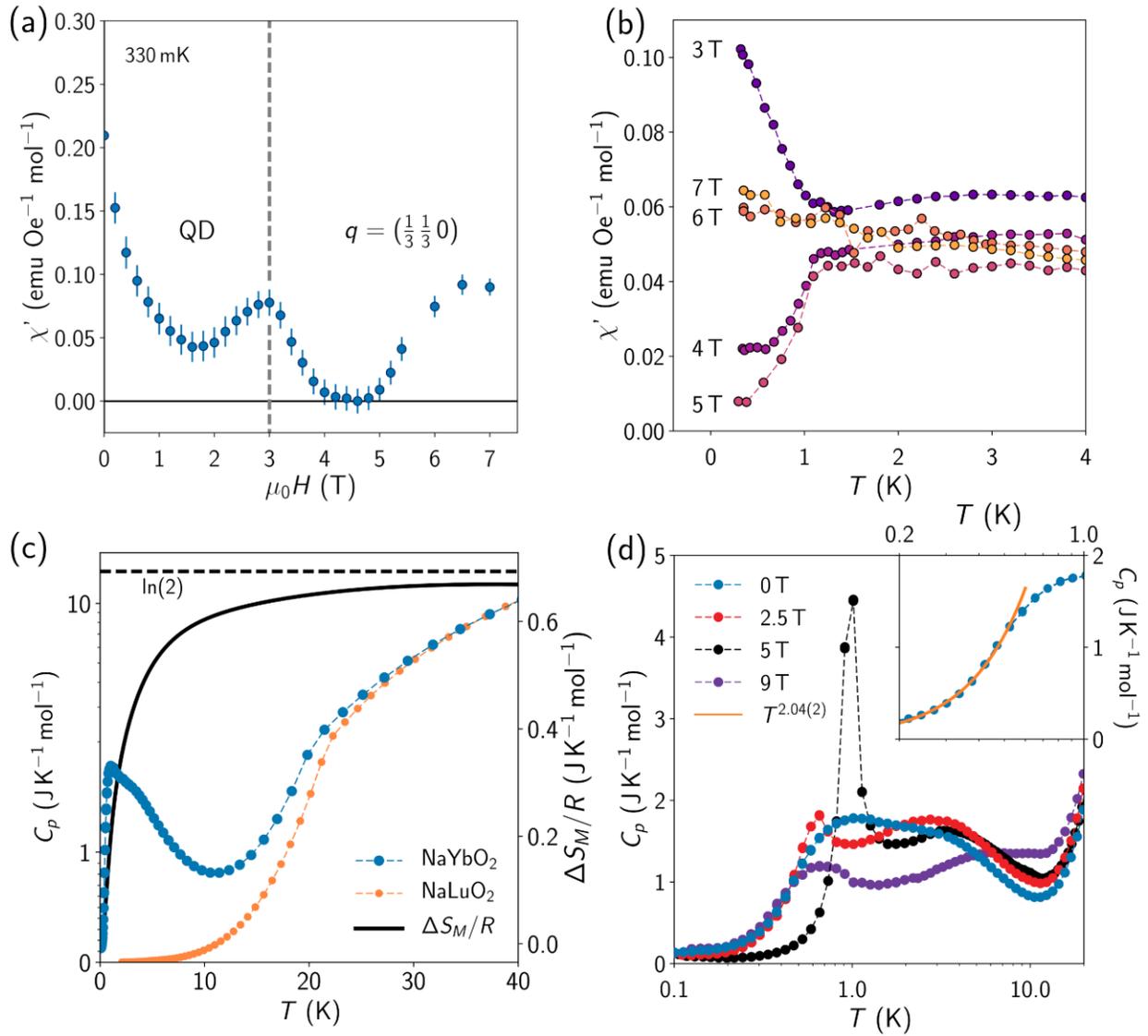

**Fig. 3: High-field magnetic susceptibility and heat capacity data.** (a) $\chi'(H)$ data collected at 330 mK showing the phase boundary between the quantum disordered (QD) ground state and magnetically-ordered **q**=(1/3, 1/3, 0) state near 3T. A second transition back into the QD state or a quantum paramagnetic phase begins to onset at higher fields. (b) $\chi'(T)$ data collected under a series of magnetic fields that traverse the ordered state. 4T and 5T $\chi'(T)$ data illustrate the onset of the ordered phase below 1K, while 6T and 7T data suggest partial reentry into a disordered magnetic state. (c) Specific heat of NaYbO$_2$ measured down to 80 mK under zero field and overpotted with the nonmagnetic NaLuO$_2$ analogue. The resulting magnetic entropy approaches 95% of Rln(2). (d) Specific heat as a function of temperature C$_p$(T) under varying magnetic fields. The lower peak centered around 1K develops a sharp anomaly at 5T indicative of the phase transition into the **q**=(1/3, 1/3, 0) state that is suppressed by 9T. Error bars denote one standard deviation of the data.

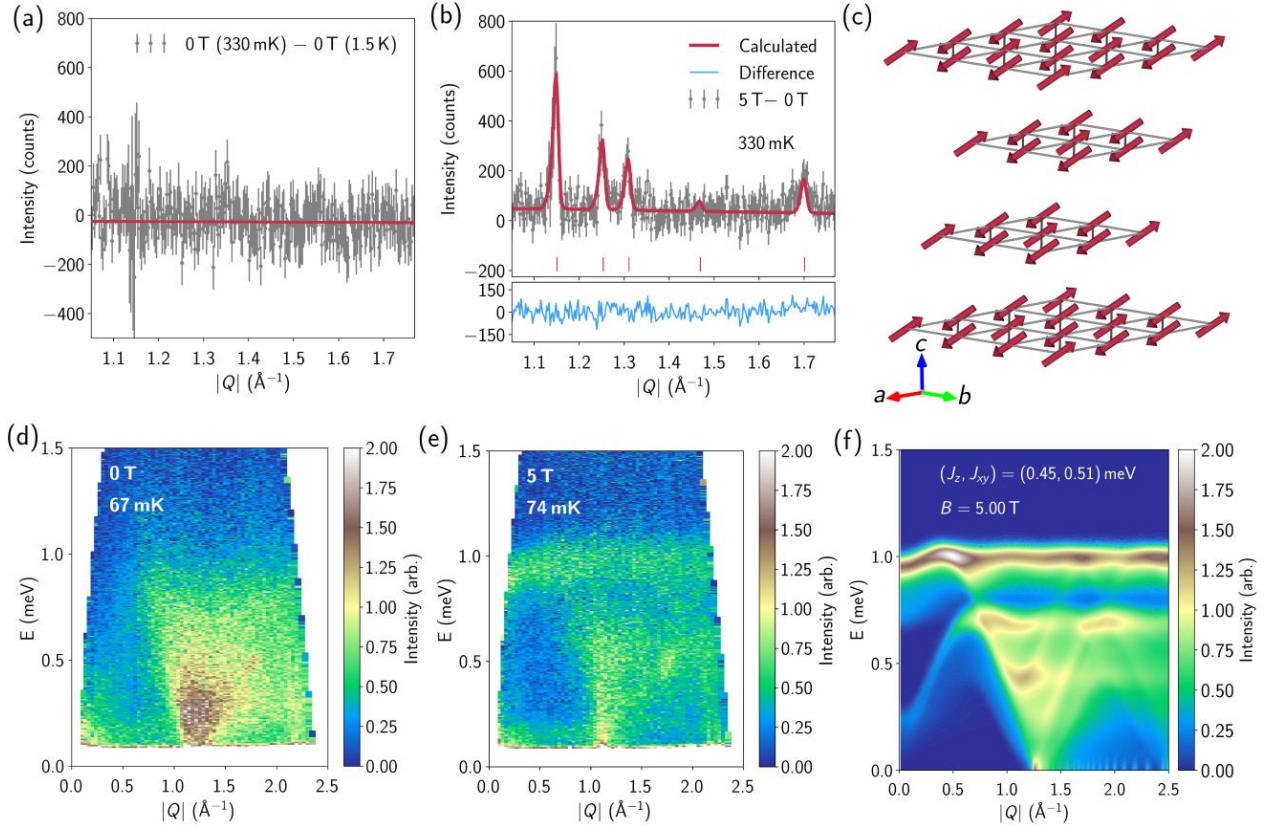

**Fig. 4: Neutron diffraction and inelastic neutron scattering data.** (a) Temperature subtracted neutron powder diffraction data (330mK – 1.5K) collected under 0T showing the absence of low temperature magnetic order. (b) Under an applied field of 5T at 450mK, new magnetic peaks appear at (1/3, 1/3, z) positions (z = 0, 1, 3), corresponding to an ordering wave vector of **q**=(1/3, 1/3, 0). The data were refined by analyzing field subtracted data (5T – 0T) which is constrained by the suppressed (1/3, 1/3, 2) reflection. (c) The best fit to the 5T induced magnetic state using the two-**q** structure **q**=(1/3, 1/3, 0) + **q**=(0, 0, 0) is generated by a collinear, modulated spin structure with Yb moments of 1.36(10) $\mu_B$. The displayed structure aligns moments approximately along the < 1, -1, -1 > direction and has six symmetrically equivalent structures generated by three-fold in-plane rotational and mirror symmetries. (d) Inelastic neutron scattering spectrum collected at 67 mK and 0T. (e) Inelastic neutron scattering spectrum collected at 74 mK and 5T (f) Linear spin wave calculations showing the powder averaged S(Q,ω) for a two dimensional triangular lattice of NaYbO$_2$'s anisotropic Yb$^{3+}$ moments in a 5T field and three-sublattice ordering. Error bars denote one standard deviation of the data.